\documentclass[pra,amsmath,amssymb,showpacs,superscriptaddress, twocolumn]{revtex4}

\usepackage{graphicx}
\usepackage{dcolumn}
\usepackage[hypertex]{hyperref}
\usepackage{bm}
\usepackage[usenames,dvipsnames]{color}
\usepackage{amsmath}
\usepackage{amsfonts}
\usepackage{amsthm}
\usepackage{amssymb}
\usepackage{mathrsfs}
\usepackage{subfigure}
\usepackage{ulem}

\begin{document}

\title{Approaching quantum-limited amplification with large gain catalyzed by hybrid nonlinear media in cavity optomechanics}

\author{Qiang Zheng}
\affiliation{Beijing Computational Science Research Center, Beijing, 100084, China}
\affiliation{School of Mathematics, Guizhou Normal University, Guiyang, 550001, China}

\author{Kai Li}
\affiliation{Beijing Computational Science Research Center, Beijing, 100084, China}

\author{Yong Li}
%\email{yongli@csrc.ac.cn}
\affiliation{Beijing Computational Science Research Center, Beijing, 100084, China}
\affiliation{Synergetic Innovation Center of Quantum Information and Quantum Physics, University of Science and Technology of China,
Hefei, Anhui 230026, China}

\date{\today}

\begin{abstract}
Amplifier is at the heart of almost all experiment carrying out the precise measurement of a weak signal.
An idea amplifier
should have large gain and minimum added noise simultaneously. Here, we consider
the quantum measurement properties of a hybrid nonlinear cavity
with the Kerr and OPA nonlinear media to amplify an input signal. We show that our hybrid-nonlinear-cavity
amplifier has large gain in the single-value stable regime and achieves quantum limit unconditionally.
\end{abstract}


\maketitle

\section{INTRODUCTION}
According to the basic rule of quantum mechanics, any amplifier always spoils the input signal by adding
a certain amount of noise \cite{Devoret2000}. The added noise for a linear phase-preserving amplifier is at least the half quantum of the signal \cite{caves1982}. The non-degenerate parametric amplifier \cite{Louisell1961}, based on the coherent interaction between the signal and idler modes driven by the pump field, is the standard paradigm of quantum limit amplifier
\cite{Glauber1967}. Utilizing the experimental achievement in realizing the
non-degenerate parametric amplifier by superconducting circuits
\cite{Yurke1988, Lehnert2008, Yamamoto2008, Devoret2010, Devoret2010b, Siddiqi2011}, it is used in turn to measure mechanical
motion near quantum limit \cite{Siddiqi2009}, probe the quantum jump of a superconducting qubit
\cite{Siddiqi2011a, Hatridge13}, and stabilize quantum coherence \cite{Siddiqi12a},
etc. Moreover, other kind of amplifiers, such as the traveling-wave parametric amplifiers \cite{Eom12}, which
do not need the cavity but require phase-matching, and the probabilistic amplifiers \cite{caves2013}, which can amplify the signal noiselessly, have also excited the wide interest.

Cavity optomechanics \cite{Aspelmeyer2012a, Aspelmeyer2014a}, in which a mechanical oscillator couples to an optical field by radiation pressure, is another active realm in recent years. Many interesting progresses have been achieved, such as ground state cooling of the mechanical resonator \cite{Rae2007}, optomechanically induced transparency \cite{Agarwal2010, Weis2010, Painter2011}, optomechanical entanglement \cite{LTYDWANG2013, Barzanjeh2012a, Teufel2013} and EPR steering \cite{QYH2013}, optomechanical squeezing \cite{Kronwald2013, Wollman2015, Agarwal2015}, just to name a few. Cavity optomechanics has standout advantage to explore the quantum behavior at macroscopic level
\cite{Meystre2012}, as well as quantum information processing \cite{Mancini2003, Stannigel2011}, quantum-classical transition
\cite{Marshall2003, cpsun2007}, quantum illumination \cite{shbar15}, and quantum heat engine \cite{Zhang2014a}.

Except for these promising advances, cavity optomechanics has also been used to realize quantum
limited amplifier. A standard radiation-pressure-coupling optomechanical system can amplify the
input signal with large gain near the quantum limit if the cavity is driven in the blue-detuned regime \cite{Massel2011}.
Some other interesting schemes, such as the
reservoir engineering \cite{Metelmann2014, Metelmann2014b, xylu2015} and the reversed dissipation \cite{Nunnenkamp2014}, have been proposed to achieve large gain quantum-limited amplifier. All the above mentioned cavity amplifiers are just the scattering mode of operation. In addition to the scattering mode, the so-called operational-amplifying (op-amp) mode \cite{Clerk2011} is another kind of amplifier. The back-action originating from the interaction between the signal and amplifier, which is
absent in the scattering mode, is essential to determine the quantum limit of the op-amp mode amplifier \cite{Clerk2011}.

Usually, using the linearly driven cavity as the op-amp amplifier, the position of
a mechanical resonator had been measured  near the quantum limit in the well-known experiment \cite{Schwab2010b}. It is also an interesting question to explore the performance of nonlinear-cavity amplifier. Actually, the nonlinear cavity as the op-amp mode amplifier has been used to measure the qubit with large gain in quantum limit \cite{Ong2011, Khan2014}.
With the Kerr nonlinear medium in the cavity, the quantum limit of the cavity amplifier in the
op-amp mode has been studied \cite{clerk2011}. To obtain the large gain, the amplifier operates near (but below) the single-bistability bifurcation point \cite{clerk2011} as close as possible. We would like to stress that this operating condition is not very robust. It is known that only a very small perturbation of system parameter or external noise may make the system randomly jump up and down in the hysteresis curve.

In this paper, we suggest a hybrid-nonlinear-cavity amplifier, which is composed of a Kerr medium and a degenerate optical parametric amplifier (OPA) \cite{Agarwal2006c} nonlinear medium in the cavity, as displayed in Fig.~\ref{systema}, to overcome the drawback of Kerr-nonlinear-cavity amplifier. We find that: (i) our hybrid nonlinear cavity amplifier can have large gain in the single-value stable regime, (ii) the hybridity of Kerr and OPA is crucial
to achieve the large gain, and (iii) the nonlinear cavity amplifier operates at the quantum limit unconditionally in the low signal frequency.

The structure of this paper is as follows.
In Sec. III, we investigate the steady state and its stability of
our hybrid nonlinear-cavity amplifier, and find in the single-value stable regime, the cavity amplifier has a large gain. Next, in Sec. IV we explore whether our cavity amplifier
can approach the quantum limit in low signal frequency regime. Finally, a conclusion is provided in the last section. The basic properties of quantum limit of the op-amp mode amplifier are reviewed in Appendix.

\section{Hybrid nonlinear cavity amplifier}
\begin{figure}[!tb]
\centering
\includegraphics[width=3.0in]{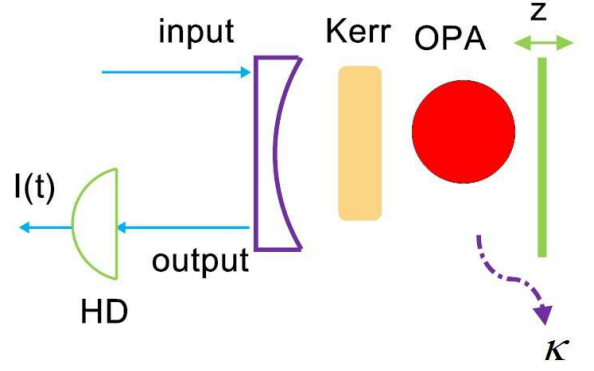}
\hspace{2.2cm}\caption{ The schematic of the cavity amplifier with the
hybrid nonlinear crystal, composed of the Kerr and OPA media.
Here $\hat{z}$ is the input signal to be
probed. The output field from the cavity is measured by the homodyne detection(HD), and the measurement current $\hat{I}(t)$ is produced. } \label{systema}
\end{figure}

We firstly consider the properties of the cavity amplifier, which has the hybrid nonlinear
crystal composed of a degenerate OPA and Kerr mediums.
The cavity with the resonant frequency $\omega_{c}$ and the decay rate $\kappa$ is also strongly driven by the input laser with frequency $\omega_{d}$, as shown in Fig.~\ref{systema}. In the rotating frame, the Hamiltonian of the cavity amplifier can be written as ($\hbar=1$ in the following)
\begin{equation}
\hat{H}_{sys}=-\Delta \hat{a}^{\dag} \hat{a}+i G(e^{i \theta }\hat{a}^{\dag 2}-h.c. )-
\Lambda \hat{a}^{\dag 2} \hat{a}^2 +i \varepsilon ( \hat{a}^{\dag}- \hat{a}).
\label{hamA}
\end{equation}
Here $\hat{a}$ ($\hat{a}^{\dag}$) is the annihilator (creation) operator of the cavity field. $\Delta=\omega_{d}-\omega_{c}$ is the detuning between the cavity field and the driving laser. $G$ is the nonlinear gain coefficient of the OPA, which is proportional to the strength of the coherent pump field driving the OPA, and $\theta$ is the phase of the OPA pumping field. $\Lambda$ and $\varepsilon$ are the Kerr coefficient and the strength of the cavity driving laser, respectively.

The first term in Eq.~(\ref{hamA}) denotes the energy of the cavity field. The
second and third terms arise from the coupling between the OPA or Kerr medium and the cavity field, respectively, and the last term is the interaction between the cavity mode and the input driving laser.

Previously, the optomechanically induced transparency and the
cooling of the mechanical resonator with this kind of hybrid-nonlinear cavity had been investigated \cite{Shahidani2013, Shahidani2014}.
Very recently, the cavity with only the OPA medium has been suggested to enhance quantum-limited position detection \cite{Marquardt2015b}. And an amplified interferometer based on the OPA medium can be used to
detect the state of a qubit \cite{Barzanjeh2014}. Note that the amplifiers proposed in Refs. \cite{Marquardt2015b, Barzanjeh2014} are operated in the scattering mode. In contrast, this paper explores the
performance of this hybrid-nonlinear cavity amplifier in the op-amp mode.

\subsection{Steady state of cavity amplifier and its stability}
Taking the cavity decay into consideration, we adopt the well-known input-output theory
\cite{walls2008}. In this frame, the motion of the cavity field can be described by the
Heisenberg Langevin equation, which is derived as
\begin{equation}
\frac{d \hat{a}}{ d t}=i \Delta \hat{a}+2G e^{i \theta }\hat{a}^{\dag} +2i \Lambda \hat{a}^{\dag} \hat{a}^{2}+\varepsilon-\frac{\kappa}{2} \hat{a}-\sqrt{\kappa} \hat{\xi}.
\label{moe0A}
\end{equation}
Here $\hat{\xi}$ is quantum vacuum fluctuation with zero mean. With the strong external driving laser, the cavity mode can be decomposed into
\begin{equation}
\hat{a}=\alpha +\hat{d}
\end{equation}
with the coherent part $\alpha=\langle \hat{a} \rangle$ and a small quantum fluctuation $\hat{d}$. According to Eq.~(\ref{moe0A}), the coherent part $\alpha$ in the steady state is determined by
\begin{equation}
\alpha= 2\varepsilon \frac{4   G + 2 i \Delta + \kappa +4 i \Lambda \bar{n} }{\kappa^2 -16 G^2+ 4 (\Delta + 2 \bar{n} \Lambda)^2 }
\label{alphaA1}
\end{equation}
with $\bar{n}=|\alpha|^2$ being the mean photon number $\hat{n}= \hat{a}^{\dag} \hat{a}$ in the cavity. It's easy to obtain the fifth-order equation for $\bar{n}$
\begin{equation}
A_{5}\bar{n}^{5}+A_{4}\bar{n}^{4}+A_{3}\bar{n}^{3}+A_{2}\bar{n}^{2}+A_{1}\bar{n}+A_{0}=0,
\label{numEq}
\end{equation}
where the coefficients are given as $A_{5}= 256 \Lambda^4$, $A_{4}= 512 \Delta \Lambda^3$,
$A_{3}= (384\Delta^2+32 \kappa^2 -512G^2 )\Lambda^2$,
$A_{2}= ( 128 \Delta^3 + 32 \Delta \kappa^2 -64 \varepsilon^2 \Lambda-512 G^2 \Delta)\Lambda$,
$A_{1}=(4 \Delta^2 + \kappa^2-16 G^2)^2 - 64 \varepsilon^2 \Delta \Lambda$, and
$A_{0}=- 4 \varepsilon^2(4 \Delta^2 + (4 G + \kappa)^2)$
for $\theta=0$.

Mathematically, the fifth-order equation, which can not be solved analytically,
should have five solutions, in which at most three roots would be stable.
For the cavity amplifier being in the stable operation condition, we should
avoid the multi-stability for the fifth-order equation. In the following,
we will solve the fifth-order equation Eq.~(\ref{numEq}) numerically, and choose the parameters to make sure
that Eq.~(\ref{numEq}) has a single real root, denoted it as $n_{s}\equiv \bar{n}$.

In general, the coherent part $\alpha$ is complex. For simplicity, it is possible to choose
the Kerr coefficient as $\Lambda=\Lambda_{0}\equiv -\Delta(4 G - \kappa )^2/( 8 \varepsilon^2)$ such that
$\alpha$ being real. Here, we restrict our study to this situation. Usually, the photon blockade induced by the Kerr interaction can greatly suppress the photon number in the cavity. However,
this mechanism does not work in our case. Reexpressing the Kerr interaction as $\hat{H}_{kerr}= \Lambda_{0} \hat{n} (\hat{n}-1)$ with very small value $\Lambda_{0} \ll 1$, it can produces a self-determined phase to ensure $\alpha$ being real. Usually, one can adjust the phase of the external driving laser to make $\alpha_{s}$ being real. In this case we will get a seventh-order equation for $n_{s}$. To simplify, in this paper we adopt the scheme by choosing appropriate Kerr coefficient.

Introducing the quadrature operators for the quantum fluctuation of the cavity amplifier
\begin{equation}
\begin{array}{llll}
\hat{x} = \frac{1}{\sqrt{2}}(\hat{d}+ \hat{d}^{\dag}),~~
\hat{p} = \frac{i}{\sqrt{2}}(\hat{d}^{\dag}-\hat{d} ),
\end{array}
\end{equation}
and similarly for $\hat{x}_{in}$ and $\hat{p}_{in}$,
with the standard linearizing approximation, i.e., $n_{s} \gg 1$,
the equation of motion for the quantum fluctuation can be rewritten as
\begin{subequations}
\begin{eqnarray}
\frac{d \hat{x}}{dt}= m_{11} \hat{x}+m_{12} \hat{p}-\sqrt{\kappa} \hat{x}_{in}
\\
\frac{d \hat{p}}{dt}= m_{21} \hat{x}+m_{22} \hat{p}-\sqrt{\kappa} \hat{p}_{in}
\end{eqnarray}
\label{moeA}
\end{subequations}
with $m_{11}=-\kappa/2+2G \cos{\theta}$,~$m_{12}= -(\Delta+ 2 \Lambda n_{s}-2G \sin{\theta})$,~
$m_{21}= \Delta+6 \Lambda n_{s}+2G \sin{\theta}$, and $m_{22}=-(\kappa/2+2G \cos{\theta})$.
The eigenvalues of the matrix
\begin{equation}
\mathbf{M}=\left(
\begin{array}{cc}
m_{11} & m_{12}\\
m_{21} & m_{22}\\
\end{array}%
\right)
\end{equation}
are obtained as
\begin{equation}
\lambda_{1,~2}=\frac{1}{2}(m_{11}+m_{22})\pm \sqrt{K},
\end{equation}
with $K=(m_{11}-m_{22})^2+4 m_{12}m_{21}$. The stability of the cavity amplifier requires that
all the eigenvalues $\lambda_{1,~2}$ must have the negative real parts.
Thus, for a positive number $K>0$ with the condition
\begin{equation}
\lambda_{1}=m_{11}+m_{22}+ \sqrt{K}=-\kappa + \sqrt{K}<0,
\label{eigst}
\end{equation}
we can sustain the stability of the cavity amplifier.
That implies the cavity decay $\kappa$ should be large enough.
Physically, this can be understood as following.
The OPA as the gain medium can magnify the photon number in the cavity, which may make it unstable.
To make sure that the cavity amplifier works in the stability regime, a large decay rate of the cavity is necessary to suppress the amplification of the photon number in the cavity.

\subsection{Gain of cavity amplifier}
From the input-output relation
\begin{equation}
\hat{x}_{out}=\sqrt{\kappa} \hat{x}+\hat{x}_{in},
\label{outputX}
\end{equation}
by solving Eq.~(\ref{moeA}) in the frequency domain, the $\hat{x}$ quadrature of the optical field can be expressed as
\begin{equation}
\hat{x}_{out}[\omega] = \tilde{g}_{x}[\omega] \hat{x}_{in}+ \tilde{g}_{p}[\omega] \hat{p}_{in},
\end{equation}
where
\begin{equation}
\tilde{g}_{x}[\omega] = 1+  \kappa (i\omega-m_{22} )/J[\omega],~~
\tilde{g}_{p}[\omega] =\kappa m_{12}/J[\omega],
\end{equation}
and
\begin{equation}
J[\omega]= m_{12}m_{21}+( im_{11}+\omega )( im_{22}+\omega).
\end{equation}
Through this paper, the Fourier transformation
is defined as $\hat{C}[\omega]=\int dt \hat{C}(t) e^{-i\omega t}$.

\begin{figure}[!tb]
\centering
\includegraphics[width=2.5in]{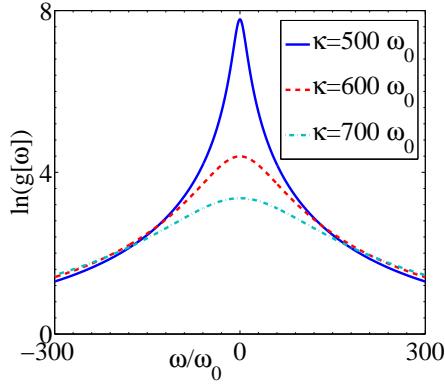}
\hspace{2.2cm}\caption{ The gain $g[\omega]$ as the function of the signal frequency $\omega$. The other parameter is  $G=120 \omega_{0}$, $\varepsilon=10^3\omega_{0}$, $\Delta=-10\omega_{0}$. $\omega_{0}$ is a
parameter with the dimension of frequency. The stability of the cavity amplifier is checked numerically with these parameters. } \label{gainA}
\end{figure}

We define the gain of the $\hat{x}$ quadrature as
\begin{equation}
g[\omega]\equiv |\tilde{g}_{x}[\omega]|^2.
\label{gainA0}
\end{equation}
In Fig.~\ref{gainA}, we plot the variation of the gain as a function of the frequency.
It displays that with the increase of the cavity decay,
the maximum value of the gain decreases considerably.
This figure also shows the gain-bandwidth tradeoff of an amplifier: with the enhanced
gain, the signal bandwidth also goes down.

From Fig.~~\ref{gainA}, it is apparent that the largest gain is obtained near $\omega=0$ with a finite
bandwidth $B_{\omega}$. Assuming that the input signal in a very narrow band $\omega \ll B_{\omega}$ with large gain, we can ensure that the amplifier operates in the zero frequency limit $\omega \rightarrow 0$. In the following, we only focus on this limit.

\begin{figure}[!tb]
\centering
\includegraphics[width=2.5in]{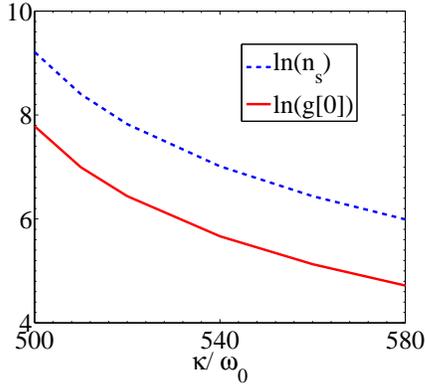}
\hspace{2.2cm}\caption{ The variation of the mean photon number $n_{s}$ and the zero
frequency gain $g[0]$ with respect to the cavity decay $\kappa$. The other parameter is the same as Fig.~\ref{gainA}. } \label{gainB}
\end{figure}

Fig.~~\ref{gainB} shows that the zero-frequency gain $g[0]$ goes down with the increase of the cavity decay rate. As the OPA is the amplification medium, it is important to explore the effect of OPA on the performance
of the cavity amplifier. We find that $g[0]$ is very sensitive to the value of OPA coefficient $G$. For example, the value of $g[0]$ goes down from $2.4 \times 10^3$ to $ 1.2 \times 10^3$ with the small change of the value of $G$ from $G=120$ to $G=118$.

To obtain a large gain for the amplifier, Ref.~\cite{clerk2011} suggest that it operates
approaching (but below) the single-bistability transition point for the cavity mean photon number as close as possible. However, near this transition point, the amplifier should lose its robustness very easily. That is, a very small random perturbation should induce it stepping into the bistability regime, and jump up and down randomly in the hysteresis curve. Compared with Ref.~\cite{clerk2011}, our cavity amplifier has very large amplification gain in a stable operating condition.

\section{Quantum limit of cavity amplifier}

In previous section, we investigate the amplification property of the cavity. We find that the cavity
amplifier has large gain with the aid of hybrid nonlinear crystal composed of the Kerr and OPA media. In this section, we will explore
whether this cavity amplifier can operate quantum-limited in low signal frequency limit.

\subsection{Coupling to signal}
To completely describe the performance of the cavity amplifier, we should obtain
the forward gain $\chi_{IF}$, which characterizes the effect of the input signal
on the output current of the homodyne detection. Usually, the forward gain $\chi_{IF}$ is determined by Kubo formula. Here, we can obtain it by rederiving the equations of motion for the cavity
amplifier with the signal coupled.

Under the standard linearization approximation, the coupling between the signal and the cavity amplifier $H_{int}$ in Eq.~(\ref{sigampcoup}) translates into
\begin{equation}
\hat{H}_{int}\simeq A \sqrt{n_{s}}(\hat{d}+\hat{d}^{\dag})\hat{z}\equiv \hat{F}\hat{z}
\end{equation}
with the linearized force operator $\hat{F}= A \sqrt{2 n_{s}} \hat{x}$. For this interaction Hamiltonian,
the equation of motion for the cavity field becomes
\begin{subequations}
\begin{eqnarray}
\frac{d \hat{x}}{dt} &=& m_{11} \hat{x}+m_{12} \hat{p}-\sqrt{\kappa} \hat{x}_{in},
\\
\frac{d \hat{p}}{dt} &=& m_{21}\hat{x}+m_{22} \hat{p} + A\sqrt{2n_{s}} \hat{z}-\sqrt{\kappa} \hat{p}_{in}.
\end{eqnarray}
\label{moeB}
\end{subequations}
Compared with Eq.~(\ref{moeA}), the equation for the operator $\hat{p}$
is changed under the effect of this coupling. For the homodyne detection, the current operator is defined as
\begin{equation}
\hat{I}_{out}=\sqrt{\kappa} ( \cos{\phi_{h}}\hat{x}_{out}+ \sin{\phi_{h}}\hat{p}_{out} ).
\end{equation}
Here $\hat{x}_{out}$ and $\hat{p}_{out}$, determined by Eq.~(\ref{moeB}), are the corresponding output quadratures of the cavity field. According to Eq.~(\ref{fowgain}), by solving Eqs.~(\ref{moeB}) in the frequency domain, the forward gain of the amplifier is obtained as
\begin{equation}
\chi_{IF}[0]= \frac{ A \kappa\sqrt{2n_{s}} }{J[0]}(m_{11}\sin{\phi_{h}}-m_{12}\cos{\phi_{h}}).
\end{equation}
Here we only focus on the zero frequency limit. This expression displays that the forward gain
is proportional to the cavity decay rate. It is reasonable physically: the cavity, as the role of a detector,
with large decay can make it respond to the input signal very quickly.

\subsection{The symmetry noise spectral density}
Without the coupling to the input signal, the homodyne current can be rewritten as
\begin{equation}
\hat{I}_{0}[\omega]= f_{1}[\omega] \hat{x}_{in}+f_{2}[\omega] \hat{p}_{in}.
\end{equation}
Here the two functions $f_{1}[\omega]$ and $f_{2}[\omega]$ are derived from Eq.~(\ref{moeA}) with the Fourier transformation. In the zero frequency limit, they are expressed as
\begin{equation}
f_{1}[0] = \sqrt{\kappa} \Gamma_{1}/J[0],
~~~f_{2}[0] = \sqrt{\kappa} \Gamma_{2}/J[0],
\end{equation}
with
\begin{equation}
\nu_{1}=J[0] - \kappa m_{22},
\\
\nu_{2}=J[0] - \kappa m_{11},
\end{equation}
and
\begin{subequations}
\begin{eqnarray}
 \Gamma_{1}=\nu_{1} \cos{\phi_{h}} + \kappa m_{21 }\sin{\phi_{h}},
 \\
 \Gamma_{2}=\nu_{2} \sin{ \phi_{h}} + \kappa m_{12} \cos{\phi_{h}}.
\end{eqnarray}
\end{subequations}

Using the correlations between the noises \cite{pzoller00}
\begin{subequations}
\begin{eqnarray}
\langle x_{in}[\omega]x_{in}[\omega'] \rangle=\langle p_{in}[\omega]p_{in}[\omega'] \rangle=\pi \delta(\omega+\omega'),
\\
\langle x_{in}[\omega]p_{in}[\omega'] \rangle=-\langle p_{in}[\omega]x_{in}[\omega'] \rangle=i \pi \delta(\omega+\omega'),
\end{eqnarray}
\end{subequations}
the symmetry noise spectral density of the homodyne current is obtained as
\begin{equation}
\bar{S}_{I_{0}I_{0}}[0]= \kappa ( \Gamma_{1}^2  + \Gamma_{2}^2  )/ J[0]^2.
\label{corii}
\end{equation}

Substituting Eq.~(\ref{corii}) into Eqs.~(\ref{spezz}), the noise spectrum density of the
imprecision noise $\bar{S}_{zz}[0]$ is obtained as
\begin{equation}
\bar{S}_{zz}[0]=\bar{S}_{I_{0}I_{0}}[0]/ |\chi_{IF}[0]|^2.
\label{corzz}
\end{equation}

For the linearized back-action force $\hat{F}=A \sqrt{2 n_{s}}\hat{x}$, it can also be reexpressed as
\begin{equation}
\hat{F}[\omega]= h_{1}[\omega] \hat{x}_{in}+h_{2}[\omega] \hat{p}_{in}.
\end{equation}
In the limit $w\rightarrow 0$,
\begin{subequations}
\begin{eqnarray}
h_{1}[0] &=& -A \sqrt{2 n_{s} \kappa} m_{22}/J[0], \\
h_{2}[0] &=&  A \sqrt{2 n_{s} \kappa} m_{12}/J[0].
\end{eqnarray}
\end{subequations}
Similarly, the corresponding symmetry noise spectrum of the back-action force is given as
\begin{equation}
\bar{S}_{FF}[0]= A^2 (m_{12}^2+m_{22}^2)n_{s}\kappa / J[0]^2.
\label{corff}
\end{equation}
in the low frequency limit.

\subsection{Quantum limit of cavity amplifier}
To make the op-amp mode cavity amplifier work near the quantum limit, we must consider the
correlation between the imprecision noise and the back-action force.
Previously, the utility of using this correlations is an well-known idea
in the gravitational wave field \cite{ychen2001a}. By making use of Eq.~(\ref{speczf}),
after some extensive but straightforward calculations, the correlation between the current and the back-action force in the limit $w\rightarrow 0$ is obtained as
\begin{equation}
\bar{S}_{zF}[0]= \frac{ \mu_{1} \cos{\phi_{h}}+  \mu_{2} \sin{\phi_{h}} }
{  2 J[0] ( m_{12}\cos{\phi_{h}}-m_{11} \sin{\phi_{h}} )}
\label{corzf}
\end{equation}
with
$\mu_{1}=m_{22}J[0]+ \kappa (m_{22}^2+ m_{12}^2 )$, and $\mu_{2}=\kappa( m_{12}m_{11} +m_{22}m_{21} )-m_{12} J[0]$.

Substituting Eqs.~(\ref{corzz}), (\ref{corff}) and (\ref{corzf}) into the left-hand side of Eq.~(\ref{qunlimA}), straightforward calculation yields
\begin{equation}
\bar{S}_{zz}[0]\bar{S}_{FF}[0]- (\bar{S}_{zF}[0])^2
=\frac{1}{4} [\frac{(m_{12} \cos{\phi_{h}} + (\kappa+m_{22} ) \sin{\phi_{h}}) }{ (m_{12} \cos{\phi_{h}}-m_{11} \sin{\phi_{h}})}]^2.
\end{equation}
It's easy to see
\begin{equation}
\bar{S}_{zz}[0]\bar{S}_{FF}[0]- (\bar{S}_{zF}[0])^2 \equiv \frac{1}{4}.
\end{equation}
That implies the cavity amplifier unconditionally works in the quantum limit.
To obtain this result, we use $-m_{11}= \kappa+m_{22}$.

Last but not least, for a quantum limited amplifier, the signal susceptibility $\chi_{zz}[\omega]$
should satisfy the following two optimal conditions \cite{clerk2003a}
\begin{subequations}
\begin{eqnarray}
|\chi_{zz}|&=& \sqrt{ \bar{S}_{zz}[\omega]/\bar{S}_{FF}[\omega] },
\label{sigalsuspA}
\\
\frac{\mathrm{Re} \chi_{zz} }{|\chi_{zz}|} &=& \frac{\mathrm{Re} \bar{S}_{zF}[\omega]}{\bar{S}_{zz}[\omega] \bar{S}_{FF}[\omega] }
\label{sigalsuspB}
\end{eqnarray}
\end{subequations}
in order to saturate the inequality Eq.~(\ref{ampboudA0}). Usually, the condition Eq.~(\ref{sigalsuspA})
can be satisfied by tuning of the coupling strength $A$ between the signal and detector. And
the condition Eq.~(\ref{sigalsuspB}) implies that $\mathrm{Im} \chi_{zz}[\omega]=0$, which can be achieved for a
damped harmonic oscillator working far from the resonant. In this paper, we assume that the conditions in Eqs.~(\ref{sigalsuspA}) and Eqs.~(\ref{sigalsuspB}) are achieved and our cavity amplifier works in the
the zero signal frequency limit, thus we mainly
concentrate on the quantum limit inequality Eq.~(\ref{qunlimA}).

\section{Conclusion}
In summary, in the system that the nonlinear cavity with a degenerate OPA and Kerr media,
we demonstrate that as the op-amp mode of operation, the cavity amplifier has the large gain. We
also find that the added noise by the nonlinear cavity amplifier is quantum-limited.
Distinct from only the Kerr medium in the cavity, which realizes the
large gain by operating near the single-bistable transition, the cavity amplifier with
OPA and Kerr nonlinear mediums can have large gain in single-value stable regime. As this
paper only investigates the low input signal frequency limit, in the future it will
be an exciting topic to explore the effect of the bandwidth on the performance of this
hybrid OPA and Kerr nonlinear cavity amplifier.
Our study may be relevant to the implement of ultrasensitive probing the
feeble signal.

\begin{acknowledgments}
We thank A. Metelmann for his helpful discussions.
This work was supported by the National Natural Science Foundation of China (Grants No. 11365006, No. 11364006
No. 11422437), the 973 program
(Grants No. 2012CB922104 and No. 2014CB921403), and the National Natural Science Foundation of Guizhou Province
QKHLHZ[2015]7767.
\end{acknowledgments}

\appendix
\section{General condition of quantum-limited amplification}
Here, we review the main aspects of the quantum limit of cavity amplifier in op-amp mode \cite{Clerk2011}. The input signal $\hat{z}$ to be probed interacts with the cavity photon number $\hat{n}= \hat{a}^{\dag}\hat{a}$ as
\begin{equation}
\hat{H}_{int}=A \hat{a}^{\dag} \hat{a} \hat{z}.
\label{sigampcoup}
\end{equation}
It is a standard radiation-pressure coupling in cavity optomechanical system, if $\hat{z}$ denotes the position of a nanomechanical oscillator. This dispersive interaction is also used to probe the state of the superconducting qubit \cite{Blais2004}. Here $A$ is a constant, and $\hat{F}=A \hat{a}^{\dag} \hat{a}$ is the back-action force. The interaction can shift the
frequency of the cavity field and the phase of the output field. To probe this phase shift, we adopt the homodyne detection, where the output from the cavity field $\hat{a}_{out}(t)$ is mixed with a reference beam. The homodyne current $\hat{I}(t)$ is described by
\begin{equation}
\hat{I}(t)=B \sqrt{\kappa/2} [e^{i\phi_{h}}\hat{a}_{out}(t)+h.c.].
\end{equation}
Here $B$ is the dimensional constant, and $\phi_{h}$ is the phase of the reference beam. Without loss of generality, we set $B=1$ throughout this paper.

Supposing that the coupling between the signal and the cavity amplifier is weak enough,
according to linear response theory \cite{Altland2006}, as a good approximation, the output current should  linearly relate to the input signal
\begin{equation}
\langle \hat{I}(t) \rangle=\langle \hat{I}_{0}(t) \rangle +\int_{-\infty}^{\infty}dt' \chi_{IF}(t-t')\langle \hat{z}(t') \rangle.
\label{fowgain}
\end{equation}
Here $\langle \hat{I}_{0}(t) \rangle$ is the current without the coupling to the signal.

From this linear relation, the imprecision noise spectral density is \textit{introduced} as
\begin{equation}
\bar{S}_{zz}[\omega] \equiv \bar{S}_{I_{0}I_{0}}[\omega]/ |\chi_{IF}[\omega]|^2
\label{spezz}
\end{equation}
in the frequency domain, which describes the \textit{effective} signal fluctuation referred from the
current noise. Here the symmetrized spectral density of output fluctuation is defined as
\begin{equation}
\bar{S}_{II}[\omega]=\frac{1}{2}\int_{-\infty}^{\infty} dt e^{i w t} \langle \{ \hat{I}(t), \hat{I}(0) \} \rangle
=\frac{1}{2}(S_{II}[\omega] +S_{II}[-\omega])
\end{equation}
with the average $\langle~\rangle $ taking with respect to the state of the uncoupled detector. The noise
spectral density is defined as
\begin{equation}
2 \pi \delta(\omega+\omega')S_{II}[\omega]=\langle \hat{I}[\omega]\hat{I}[\omega'] \rangle.
\end{equation}

Except for the imprecision noise, the back-action noise is another essential
side for the op-amp mode of operation. As a result, the total added noise produced by the cavity amplifier, is composed of the back-action noise originated from Eq.~(\ref{sigampcoup}) and the imprecision noise in Eq.~(\ref{spezz}). The total added noise can be considered as the effective temperature $T_{N}[\omega]$, which has the standard bound \cite{Clerk2011, Clerk2004b}
\begin{equation}
\frac{k_{B}T_{N}[\omega]}{\hbar w} \geq
\frac{1}{\hbar}({ \sqrt{\bar{S}_{zz}[\omega]\bar{S}_{FF}[\omega]-
(Re \bar{S}_{zF}[\omega])^2 } -Im \bar{S}_{zF}[\omega] }),
\label{ampboudA0}
\end{equation}
where $\bar{S}_{FF}[\omega]$ is the noise spectral density of the back-action force, and $\bar{S}_{zF}[\omega]$ describes the correlation between the back-action and impression noises
\begin{equation}
\bar{S}_{zF}[\omega]=\bar{S}_{IF}[\omega]/\chi_{IF}[\omega].
\label{speczf}
\end{equation}
The symmetry spectrum is defined as
\begin{equation}
\bar{S}_{IF}[\omega]=\frac{1}{2} ( S_{IF}[\omega]+ S^{*}_{IF}[-\omega] ).
\end{equation}
Note the fact that there is no the reverse gain
$\chi_{FI}=0$ has been used in deriving Eq.~(\ref{ampboudA0}).

For the simplest case that the signal frequency is much smaller than the relevant quantity,
we can consider the zero signal frequency limit $\omega \rightarrow 0$.
Under this condition, Eq.~(\ref{ampboudA0}) reduces to the Heisenberg inequality \cite{Braginsky1992}
\begin{equation}
\bar{S}_{zz}[0]\bar{S}_{FF}[0]- (\bar{S}_{zF}[0])^2 \geq \hbar^2/4.
\label{qunlimA}
\end{equation}
This corresponds to the fact that the amplifier must add noise at least the same as the zero-point fluctuation
of the signal source $k_{B} T_{N} / \hbar \omega \geq \frac{1}{2}$. We would like to stress that even though the quantum limit of the op-amp mode
resembles to the counterpart of the scattering mode, they are completely different. There is no
back-action in the scattering mode.

%%%%%%%%%%%%%%%%%%%%%%%%%%%%%%%%%%%%%%%%%%%%%%%%%%%%%%%%%%%%%%%%%%%%%%%%%%%%%%%%%%%%%%%%%%%%%

%%%%%%%%%%%%%%%%%%%%%%%%%%%%%%%%%%%%%%%%%%%%%%%%%%%%%%%%%%%%%%%%%%%%%%%%%%%%%%%%%%%%%%%%%%%%%

\end{document}